\documentclass[12pt,notitlepage,a4paper]{article}

\pdfoutput=1

\usepackage{a4wide}
\usepackage{epsfig}
\usepackage{color,graphicx}
\usepackage{xcolor}

\usepackage{amssymb,amsmath}
\usepackage{jheppub}
\usepackage{adjustbox}
\usepackage{multirow}
\usepackage{makecell}
\newcommand{\be}{\begin{equation}}
\newcommand{\ee}{\end{equation}}
\newcommand{\bea}{\begin{eqnarray}}
\newcommand{\eea}{\end{eqnarray}}
\newcommand{\nn}{\nonumber}

\begin{document}

\begin{center}  

\vskip 2cm 

\centerline{\Large { A note on dualities of F(4) type 3d $\mathcal{N}=5$ SCFTs }}

\vskip 1cm

\renewcommand{\thefootnote}{\fnsymbol{footnote}}

   \centerline{
    {\large \bf Ki-Hong Lee${}^{a}$} \footnote{khlee11812@gmail.com}, {\large \bf Belal Nazzal${}^{a}$} \footnote{sban@technion.ac.il}, {\large \bf Gabi Zafrir${}^{a,b}$} \footnote{gabi.zafrir@oranim.ac.il}}
      
\vspace{1cm}
\centerline{{\it ${}^a$ Haifa Research Center for Theoretical Physics and Astrophysics, University of Haifa,}}
\centerline{{\it Haifa 3498838, Israel}}
\centerline{{\it ${}^b$ Department of Physics, University of Haifa at Oranim, Kiryat Tivon 36006, Israel}}
\vspace{1cm}

\end{center}

\vskip 0.3 cm

\setcounter{footnote}{0}
\renewcommand{\thefootnote}{\arabic{footnote}}   
   
\begin{abstract}

 We suggest a new duality between a pair of 3d $\mathcal{N}=5$ SCFTs, one of ABJ type and one based on the exceptional superalgebra $F(4)$. Our main evidence for the proposed duality is the matching of the superconformal index. In addition to the intrinsic interest in dualities between strongly coupled field theories, the result can also be useful in the classification of 3d $\mathcal{N}=5$ SCFTs.

\end{abstract}
 
 \newpage
 
\tableofcontents
 
\section{Introduction}
\label{sec:intro}

The classification of interacting conformal field theories (CFTs) is an interesting endeavor in the study of quantum field theory. Due to the appearance of CFTs at the end of RG flows and in critical points, a classification of CFTs could be used to constrain the possible interacting low-energy theories, as well as provide a complete enumeration of the possible universality classes. However, performing a complete classification of CFTs appears to be a rather difficult feat. This problem can be simplified, while still maintaining part of its utility by restricting to CFTs possessing certain symmetries. This is especially true if one considers CFTs that are also supersymmetric, that is SCFTs. These are further constrained by the presence of superconformal symmetry, which makes SCFTs a natural playground to explore the classification of CFTs with additional properties.  

The extent and details of the classification of SCFTs depend crucially on the number of supercharges. Here we shall concentrate on SCFTs with more than $8$ supercharges, for which a classification scheme appears more manageable. These include notable cases like 6d $(2,0)$ SCFTs and 4d $\mathcal{N}=4$ SCFTs. The former are believed to be classified by a choice of $ADE$ gauge algebra, while the latter are believed to be classified by a choice of Weyl group. Inspired by these and other observations, a proposal was put forth in \cite{Tachikawa:2019dvq,Deb:2024zay} for the classification of SCFTs with more than $8$ supercharges in dimensions $3\leq d\leq 6$. The general idea was to structure the classification based on the moduli space of the SCFT. Specifically, the moduli space of SCFTs in this class generically has the form of a flat space quotiented by a discrete group $\Gamma$. The observation in \cite{Tachikawa:2019dvq,Deb:2024zay} is that for all known cases $\Gamma$ is a reflection group, with the exact type depending on $d$ and the number of supercharges\footnote{There are known cases of SCFTs in this class where $\Gamma$ is not a reflection group, but these can always be realized by gauging a discrete symmetry of an SCFT for which $\Gamma$ is a reflection group. Therefore, the proper statement of the observation is that any known SCFT with more than 8 supercharges can be realized by gauging a discrete 0-form symmetry of an SCFT for which $\Gamma$ is a reflection group.}. Most of the details would not be important to us here, and we shall merely concentrate on the details that are. We direct the interested reader to the references for further details.

The idea then is that SCFTs can be classified based on $\Gamma$. This relies on the fact that reflection groups are rather constrained, and that a complete mathematical classification exists for the relevant cases. Nevertheless, it is observed that for $3d$ $\mathcal{N}=6$ and $\mathcal{N}=5$ SCFTs (corresonding to 12 and 10 supercharges, respectively), $\Gamma$ does not uniquely determine the SCFT, that is there are different SCFTs with the same moduli space, and do correspond to the same choice of $\Gamma$ \footnote{To facilitate the classification one generally identifies theories differing by marginal operators and topological manipulations like gauging discrete symmetries. The latter specifically implies that one classifies SCFTs with no non-anomalous $d-2$ form symmetries. Since gauging a 0-form symmetry leads to a dual non-anomalous $d-2$ form symmetry, this can be used as an indication that there are no gauged 0-form symmetry, and so one expects their moduli space to be flat space quotiented by a reflection group. For 4d $\mathcal{N}=4,3$ and 3d $\mathcal{N}=8$ SCFTs these identification are sufficent for $\Gamma$ to uniquely specify the family of theories, but this fails for 3d $\mathcal{N}=6,5$ SCFTs, see \cite{Tachikawa:2019dvq,Deb:2024zay} for further details.}. A complete classification would then require us to also enumerate the number of theories corresponding to each choice. A complication here is that two seemingly different theories may describe the same low-energy theory, a phenomenon that is usually referred to as duality. As such, understanding dualities between different theories is important for the classification of SCFTs.

Dualities are also interesting in their own right, as they usually entail an intricate relation between the two theories that is often of a non-perturbative nature. Dualities also allow us to map field theory problems in one theory to potentially easier ones in the dual theory, and so can be of use in tackling QFT problems. As such, dualities enjoy considerable study, and this is also true for the case of 3d $\mathcal{N}=6$ and $\mathcal{N}=5$ SCFTs, where many dualities are known. Here we shall consider the case of $\mathcal{N}=5$ SCFTs, which are the main subject of this paper, in greater detail. Probably the most well-known family of $\mathcal{N}=5$ SCFTs are the orthosymplectic ABJ theories \cite{Aharony:2008gk,Hosomichi:2008jb}. These can be described as the low-energy limit of a stack of $M2-$branes probing a $\mathbb{C}^4/\hat{D}_k$ singularity, or alternatively as the low-energy limit of a gauge theory with $USp(2N)_k \times O(M)_{-2k}$ gauge group (subscripts denoting the Chern-Simons levels) and a bifundamental hypermultiplet. This family of theories is known to enjoy dualities relating it to itself \cite{Aharony:2008gk}, and to other theories like the unitary ABJM or ABJ theories \cite{Cheon:2012be,Beratto:2021xmn}. These dualities are then important in the precise classification of 3d $\mathcal{N}=6$ and $\mathcal{N}=5$ SCFTs.

Besides this family of theories, there are other known 3d $\mathcal{N}=5$ SCFTs. One family of theories is provided by the low-energy theory on $M2-$branes probing a $\mathbb{C}^4/\hat{E}_k$ singularity, for $k=6,7$ or $8$. These have no known Lagrangian description, complicating their study and so would not be considered here. Finally, there are three families of $\mathcal{N}=5$ SCFTs, described by certain 3d Chern-Simons matter theories. These are usually denoted by the exceptional superalgebras on which they are based: $D(2|1;\alpha)$, $F(4)$ and $G(3)$, see \cite{Gaiotto:2008sd}. These are relatively understudied, which begs the question of whether these are always unique $\mathcal{N}=5$ SCFTs or if dualities exist relating certain members among themselves or to other cases.

Our main result is evidence for a duality between a certain $\mathcal{N}=5$ SCFT based on the $F(4)$ superalgebra and an orthosymplectic ABJ theory. Specifically, the $F(4)$ SCFT in question has a Lagrangian description consisting of a $Spin(7)_{-3}\times SU(2)_{2}$ gauge group with a bifundamental hyper in the $(\bf{ 8}, \bf{ 2})$. The fact that the bifundamental is in the spinor representation of the $Spin(7)$, rather than the vector, is what differentiate this case from the ABJ theories. Nevertheless, we shall present evidence that this theory is equivalent to an ABJ type theory, that is both theories flow to the same IR $\mathcal{N}=5$ SCFT. The ABJ theory in question consists of a $USp(2)_1 \times Spin(4)_{-2}$ gauge group with a bifundamental hypermultiplet. Our main evidence supporting the duality argument is the matching of the superconformal index of the two theories.

As mentioned, the above result is useful in the classification of $\mathcal{N}=5$ SCFTs, as it is important to know how many distinct theories exist. It can also be of use for performing computations. This is especially the case, as the large rank difference in the $F(4)$ case makes the computation of various partition functions quite challenging technically. These computations, however, become more manageable on the dual ABJ side. This is also the reason why we only present a matching of the superconformal index, as the computation of other partition functions, like the round sphere one, appears to be more challenging. It would be interesting if the duality can also be checked by matching other partition functions, but we shall leave this to future work. 

Another interesting question is whether all the remaining $\mathcal{N}=5$ SCFTs based on exceptional superalgebras are distinct, or whether there are additional duality relations between them. Unfortunately, the difficulty in computing partition functions for cases based on $F(4)$ and $G(3)$ makes it difficult to check this explicitly. As such, we shall leave this for future work as well.

The structure of this paper is as follows. We begin by introducing the proposed duality in section \ref{sec:intD}. We then check the proposed duality by matching the superconformal index. The main results here are summarized in section \ref{sec:index}, with the more technical details presented in appendix \ref{app:ind}.      

\section{Introducing the duality}
\label{sec:intD}

We shall begin by introducing the theories participating in the duality. 

\subsection{The $F(4)$ $\mathcal{N}=5$ SCFT}

On one side of the duality we have an $\mathcal{N}=5$ SCFT associated with the $F(4)$ superconformal algebra. The matter content of this class of theories consists of a $Spin(7)_{-3k}\times SU(2)_{2k}$ gauge group with a bifundamental hyper in the $(\bf{ 8}, \bf{ 2})$. The moduli space of these theories is known to be $\mathbb{C}^4/\hat{D}_{6k}$ for $k>1$ and $\mathbb{C}^4/\hat{D}_{2}$ for $k=1$. As can be seen, the case of $k=1$ is special. This is also apparent by examining the residual gauge group on the moduli space. On a generic point in the moduli space the gauge group is broken to an ABJM $U(1)\times U(1)$ part and a decoupled $SU(3)_{-3k}$ part \cite{Deb:2024zay}. The special phenomena occurring at $k=1$ is that the $SU(3)_{-3}$ part is now level-rank dual to an empty theory. This is similar to the case of the $U(N+k)_k \times U(N)_{-k}$ ABJ model. Here on a generic point on the moduli space we get $N$ copies of the ABJM $U(1)\times U(1)$ part and a decoupled $U(k)_{k}$ part, which as argued in \cite{Aharony:2008gk}, the $SU(k)_k$ part of which is level-rank dual to an empty theory. This level-rank duality of the decoupled part then signals a duality of the full gauge theory, specifically between the $U(N+k)_k \times U(N)_{-k}$ ABJ model and the ABJM model $U(N)_k \times U(N)_{-k}$. It is therefore natural to ask whether the same phenomena in the $F(4)$ theory at $k=1$ is also an indicator of a duality for this SCFT. Here we shall try to argue that this is in fact correct.  

\subsection{Statement of the duality and implications}

Inspired by the above observation, we propose a duality between the following two $\mathcal{N}=5$ theories:

\be \label{dualitySpinSpin}
Spin(7)_{-3}\times SU(2)_{2} \longleftrightarrow Spin(4)_{-2}\times USp(2)_1
\ee

Here we only listed the gauge group for brevity, but there is also the bifundamental hyper, in the spinor of $Spin(7)$ for the theory on the left and the vector of $Spin(4)$ for the theory on the right, as well as a quartic superpotential required to preserve $\mathcal{N}=5$ supersymmetry (see for instance the discussion in \cite{Schnabl:2008wj}). 

Starting from this proposed duality, we can generate a new one by gauging a 1-form symmetry on both sides. Specifically, the $F(4)$ theory has a 1-form symmetry associated with the diagonal center of $Spin(7)$ and the $SU(2)$. This 1-form symmetry is non-anomalous \cite{Deb:2024zay}, and the gauging changes the gauge group to $[Spin(7)\times SU(2)]/\mathbb{Z}_2$. On the dual side, it appears that we need to gauge the 1-form symmetry acting on the spinor Wilson lines of the $Spin(4)$ gauge group, as is apparent by matching the superconformal index. This changes the gauge theory to $SO(4)_{-2}\times USp(2)_1$. This leads to the following duality:

\be
[Spin(7)_{-3}\times SU(2)_2]/\mathbb{Z}_2 \longleftrightarrow SO(4)_{-2}\times USp(2)_1
\ee 

The moduli space of both theories is known to be $\mathbb{C}^4/\hat{D}_1 = \mathbb{C}^4/\mathbb{Z}_4$, and in fact, these theories actually enjoy $\mathcal{N}=6$ supersymmetry \cite{Beratto:2021xmn,Garavaglia:2025cgz}. Furthermore, the $SO(4)_{-2}\times USp(2)_1$ theory is known to be dual to the unitary ABJ theory with gauge group $[U(3)_4 \times U(1)_4]/\mathbb{Z}_2$ \cite{Cheon:2012be}, which has manifest $\mathcal{N}=6$ supersymmetry. Overall then we get the triality:

\be
[Spin(7)_{-3}\times SU(2)_2]/\mathbb{Z}_2 \longleftrightarrow SO(4)_{-2}\times USp(2)_1 \longleftrightarrow [U(3)_4 \times U(1)_4]/\mathbb{Z}_2
\ee   

It is interesting to compare the symmetries in each of the three theories. We shall begin with the symmetry mapping between $SO(4)_{-2}\times USp(2)_1$ and $[U(3)_4 \times U(1)_4]/\mathbb{Z}_2$, where we summarize the results of \cite{Beratto:2021xmn}. Here, for convenience, we shall regard the theories as $\mathcal{N}=2$ theories. The $SO(4)_{-2}\times USp(2)_1$ theory has a continuous $SU(2)/\mathbb{Z}_2$ symmetry rotating the two bifundamental chirals. This symmetry combines with the $U(1)$ R-symmetry to form the $\mathfrak so(5)$ R-symmetry algebra of the $\mathcal{N}=5$ SCFT. Additionally, there is a $\mathbb{Z}_2$ outer automorphism symmetry, that exchanges the two spinor representations of $SO(4)$, and a $\mathbb{Z}_2$ magnetic symmetry that acts on the $SO(4)$ monopoles that are not in $Spin(4)$. Finally the theory also has a $\mathbb{Z}_2$ 1-form symmetry associated with the diagonal center of $SO(4)$ and $USp(2)$. This symmetry is known to be anomalous and so cannot be gauged.

The $[U(3)_4 \times U(1)_4]/\mathbb{Z}_2$ theory has a continuous $[SU(2)\times SU(2)]/\mathbb{Z}_2$ symmetry rotating the four bifundamental chirals. These combine with the $U(1)$ R-symmetry to form the $\mathfrak{so}(6)$ R-symmetry algebra of the $\mathcal{N}=6$ SCFT. Additionally, the theory has a $U(1)$ magnetic symmetry. These symmetries are all part of the $\mathcal{N}=6$ superconformal algebra\footnote{This follows the result of \cite{Bashkirov:2011fr}, that the energy-momentum tensor multiplet of any $\mathcal{N}=6$ SCFT also contains a conserved current for a flavor symmetry, leading to the presence an an additional $U(1)$ global symmetry.}, and are not all manifest in the $SO(4)_{-2}\times USp(2)_1$ theory as it only shows $\mathcal{N}=5$ supersymmetry. Specifically, the $U(1)$ symmetry appears to be accidental and only the diagonal $\mathfrak {su}(2)$ is manifest on the $SO-USp$ side, which is the proper embedding of the $\mathfrak{so}(5)$ R-symmetry of $\mathcal{N}=5$ susy in the $\mathfrak{so}(6)$ R-symmetry of $\mathcal{N}=6$ susy. The theory also has a discrete $\mathbb{Z}_2$ magnetic symmetry acting on the monopoles of $[U(3) \times U(1)]/\mathbb{Z}_2$ that are inconsistent in $U(3) \times U(1)$, and an anomalous $\mathbb{Z}_2$ 1-form symmetry. One can see that the 1-form symmetry matches on the two sides, while the $\mathbb{Z}_2$ 0-form symmetry of the unitary theory mapping to the $\mathbb{Z}_2$ outer automorphism symmetry of the $SO-USp$ quiver. This latter statement is evident in that the $U(3)_4 \times U(1)_4$ theory is dual to $O(4)_{-2}\times USp(2)_1$, see \cite{Beratto:2021xmn}. The $\mathbb{Z}_2$ magnetic symmetry of the $SO-USp$ theory appears to arise accidentally on the unitary side.

Now we consider the symmetries of the $F(4)/\mathbb{Z}_2$ theory. Similarly to the $SO-USp$ quiver, it only manifests $\mathcal{N}=5$ supersymmetry, and so only exhibits an $SU(2)/\mathbb{Z}_2$ continuous symmetry. It has a $\mathbb{Z}_2$ symmetry acting on the monopoles of $[Spin(7)\times SU(2)]/\mathbb{Z}_2$ that are not in $Spin(7)\times SU(2)$. The duality \eqref{dualitySpinSpin} suggests that this symmetry maps to the $\mathbb{Z}_2$ magnetic symmetry of the $SO-USp$ theory. Note that this theory has no 1-form symmetry. This suggets that the 1-form symmetry, the additional $\mathbb{Z}_2$ 0-form symmetry and the continuous symmetries required by $\mathcal{N}=6$ susy all arise accidentally.      

It is interesting that the two $\mathbb{Z}_2$ symmetries of the $SO-USp$ theory don't simultaneously appear in both dual descriptions. As such, the triality appears to only exists for the $SO(4)_{-2}\times USp(2)_1$ variant, with the $O(4)_{-2}\times USp(2)_1$ variant having a unitary dual, and the $Spin(4)_{-2}\times USp(2)_1$ variant having an $F(4)$ dual.

Having presented the two dualities we consider here, we next give supporting evidence for them using the superconformal index. 

\section{ The superconformal index}\label{sec:index}

Our main evidence for the proposed duality is the matching of the superconformal index, which we present here. We follow the conventions of \cite{Kapustin:2011jm} for the 3d superconformal index.

\subsection { Computing the indices for the dual theories}

Next we shall compute the index for the pair of theories participating in the duality. As customary, we shall expand the index in a power series in $x$ and compute up to order $x^4$%($x^3$ in the quotient case)
, as going to higher orders proved technically challenging for the $F(4)$ cases. We shall next present our results, starting with the $F(4)$ case and then moving on to the ABJ cases. 

\subsubsection{ ${F(4)}$ and ${F(4)/\mathbb{Z}_2}$}

Let's start with the SCFT based on the exceptional lie algebra $F(4)$. This is a $Spin(7)_{-3k}\times USp(2)_{2k}$ gauge theory with a hypermultiplet in the $(\boldsymbol{8},\boldsymbol{2})$ representation of the gauge group. Here we shall restrict ourselves to the $k=1$ case. We choose an $SU(2)^3\times SU(2)$ cartan subgroup of $Spin(7)\times USp(2)$ and the corresponding monopole charges are labeled by $(m_1,m_2,m_3,n)$. 

The $F(4)$ (with $k=1$) has moduli space $\mathbb{H}/\hat{D}_2$. We can consider another variant of this theory, denoted by $F(4)/\mathbb{Z}_2$, which is obtained by gauging the global one-form symmetry that the $F(4)$ possesses. In this case, the moduli space becomes $\mathbb{H}/\hat{D}_1$ for $k=1$.

We would like to compute the index for the two variants. The details can be found in Appendix \ref{f4app}. The results are the following:
\begin{align} \label{f4ind}
\mathcal{I}_{F(4)}=&1+ x + \Big(2\chi_{\boldsymbol{5} } (c)-\chi_{\boldsymbol{3} } (c) +2 \Big) x^2 + \Big( \chi_{\boldsymbol{7} }(c)-3 \chi_{\boldsymbol{3} } (c) \Big) x^3 +\\ \nonumber
&+\Big(3\chi_{\bf{9}}(c)-2\chi_{\bf{7}}(c)-\chi_{\bf5}(c)+3\chi_{\bf{3}}(c)+1\Big)x^4+\mathcal{O}(x^5)  
\end{align}

And for the $\mathbb{Z}_2$ variant:
\begin{align}\label{f4z2ind}
\mathcal{I}_{F(4)/\mathbb{Z}_2}=1&+ \Big(\chi_{\boldsymbol{3} } (c)+1\Big)x + \Big(3\chi_{\boldsymbol{5} } (c)-\chi_{\boldsymbol{3} }(c) +2\Big) x^2 
\nonumber\\
&+ \Big(3 \chi_{\boldsymbol{7} } (c)- \chi_{\boldsymbol{5}}(c)-5 \chi_{\boldsymbol{3} }(c)-1 \Big) x^3\\
&+ \Big(5\chi_\mathbf{9}(c)-3\chi_\mathbf{7}(c)-2\chi_\mathbf{5}(c)+6\chi_\mathbf{3}(c)+3\Big)x^4+ \mathcal{O}(x^5) \nonumber 
\end{align}

Here we use $\chi_{\boldsymbol{d} } (c)$ for the character of the $d$-dimensional representation of the $SU(2)/\mathbb{Z}_2$ rotating the bifundamental hyper.

Some technical notes: at each order only few monopoles give a non-zero contribution. However, in these theories the monopole ground state can go below unitarity due to fermionic zero modes, though these are not gauge invariant and so do not signal a problem with the IR behavior of the theory. All gauge invariant operators appear to be above the unitarity bound. Nevertheless, this presents technical difficulties in performing the computation, resulting in the computation being done only up to $x^4$. We note that similar technical difficulties occur in other partition functions, such as the one on $S^3$.    

 %the zero point-energy $\epsilon_0$ can be negative, our ability to compute the index to higher orders is limited, since as we go to higher and higher orders configurations with larger and larger absolute values start to contribute. In the first case we have computed the index taking into account configurations with $|\epsilon_0| \leq 11$, and for the latter we included configurations with $|\epsilon_0|\leq7$, up to some large enough GNO charges.

%Note the expansions for the two variants are quite similar, but we have additional terms in the $\mathbb{Z}_2$ variant. This is due to the fact that gauging the 1-form symmetry restricts the allowed Wilson lines and allow for additional magnetic monopoles to exist. In tables \ref{f4tab} and \ref{f4z2tab} we have listed all the contributing monopoles in both cases. 

\subsubsection{ ${SO(4)_{-2} \times USp(2)_1}$}

Now we consider an SCFT with $SO(4)_{-2k} \times USp(2)_k$ gauge group and a hypermultiplet in the bifundamental of the gauge groups. For $k=1$, this theory is known to have the same moduli space as $F(4)/\mathbb{Z}_2$, which is $\mathbb{H}/\hat{D}_1$, and we will show that indeed the superconformal indices match exactly in a power series expansion in $x$. 

Technically, this computation is simpler since it does not involve configurations with negative zero-point energy. We defer the details to Appendix \ref{SOapp}, and simply state the result:
\begin{align}\label{SOind}
\mathcal{I}_{SO(4)_{-2} \times USp(2)_1}=1&+ \Big(\chi_{\boldsymbol{3} } (c)+1\Big)x + \Big(3\chi_{\boldsymbol{5} } (c)-\chi_{\boldsymbol{3} }(c)+2 \Big) x^2 
\nonumber\\
&+ \Big(3 \chi_{\boldsymbol{7} } (c)- \chi_{\boldsymbol{5}}(c)-5 \chi_{\boldsymbol{3} }(c)-1 \Big) x^3\\
&+ \Big(5\chi_\mathbf{9}(c)-3\chi_\mathbf{7}(c)-2\chi_\mathbf{5}(c)+6\chi_\mathbf{3}(c)+3\Big)x^4+ \mathcal{O}(x^5) \nonumber 
\end{align}
This matches similar computations performed in \cite{Comi:2023lfm}.

The above result matches the index of $F(4)/\mathbb{Z}_2$, given in \eqref{f4z2ind}, and so strongly suggests that $Spin(7)_{-3}\times USp(2)_{2}/\mathbb{Z}_2 \cong SO(4)_{-2}\times USp(2)_1$.

\subsubsection{ ${Spin(4)_{-2} \times USp(2)_1}$}

Next we consider a similar SCFT but with $Spin(4)_{-2} \times USp(2)_1$ gauge group. This should be related to the previous one by gauging of a $\mathbb{Z}_2$ 1-form symmetry. As previously mentioned we expect this case to be dual to the $F(4)$ theory with $k=1$, which we can check by computing its superconformal index. The index is similar to the $SO(4)_{-2} \times USp(2)_1$ variant but now the sum of the monopoles associated with the $Spin(4)$ part is restricted to be even integer. Performing the computation results in:

\begin{align}
\mathcal{I}_{Spin(4)_{-2} \times USp(2)_1}=&1+ x + \Big(2\chi_{\boldsymbol{5} } (c)-\chi_{\boldsymbol{3} } (c)+2 \Big) x^2 + \Big( \chi_{\boldsymbol{7} }(c)-3 \chi_{\boldsymbol{3} } (c) \Big) x^3\\ \nonumber
&+\Big(3\chi_{\bf{9}}(c)-2\chi_{\bf{7}}(c)-\chi_{\bf5}(c)+3\chi_{\bf{3}}(c)+1\Big)x^4+\mathcal{O}(x^5)  
\end{align}
which matches exactly with \eqref{f4ind}. It also agrees with similar computations performed in \cite{Comi:2023lfm}. Thus, we suggest the duality $Spin(7)_{-3}\times USp(2)_{2}\cong Spin(4)_{-2}\times USp(2)_1$.

\section*{Acknowledgments}
We are grateful to Anirudh Deb for collaboration in early stages of the project. We would also like to thank Sunjin Choi, Seok Kim and Noppadol Mekareeya for helpful discussions. KHL, BN and GZ are partially supported by the Israel Science Foundation under grant no. 759/23.

\appendix

\section{ Calculations of the superconformal index}
\label{app:ind}

Here we collect further details regarding the computation of the superconformal index.
 
\subsection{ $F(4)$ and $F(4)/\mathbb{Z}_2$ } \label{f4app}

Following the definition of the superconformal index presented in \cite{Kapustin:2011jm}, one can write down the index of the $Spin(7)_{-3k} \times USp(2)_{2k}$ aka the $F(4)$ theory:

\begin{align}\label{F4ind}
\mathcal{I}_{Spin(7)_{-3k} \times USp(2)_{2k} }  = \frac{1}{96} \sum_{\{m_i\},n } \oint \prod_{i=1}^3 \frac{dz_i}{2\pi i z_i}  \frac{dy}{2\pi i y}&z_1^{-6k m_1}z_2^{-6k m_2}z_3^{-12 k m_3} y^{4k n}  x^{\epsilon_0} \nn
\\
& I_V(\{z_l,m_l\},y,n,x) I_{\chi} (\{z_l,m_l\}, y,n,x,c)
\end{align}

\noindent where $z_i,m_i$ parametrize a subgroup $SU(2)^3 \subset Spin(7)$ and their corresponding GNO monopole charges. Similary $y,n$ are the gauge fugacity and the monopole charge for the $USp(2)$ part. In the expression above the summation is over monopoles satisfying,
\be
m_1+m_2 \in \mathbb{Z},\;\;\; m_2+m_3 \in \mathbb{Z}, \;\;\; n\in\mathbb{Z}
\ee

\noindent
  The zero-point energy is given by,

\begin{align}
\epsilon_0(m_1,m_2,m_3)= & -2 |m_1|-2|m_2|-2|m_3|-2|n| -|m_1-m_2|
-|m_1+m_2|
\nn\\
&-|m_1+m_2-2m_3|-|m_1-m_2+2m_3|-|m_1-m_2-2m_3|
\nn\\
&-|m_1+m_2+2m_3|+|-m_1+m_3-n|+|m_1+m_3-n|
\nn\\
&\nn+|m_1+m_3+n|+|-m_2+m_3-n|+|m_2+m_3-n|\nn\\
&+|-m_1+m_3+n|+|-m_2+m_3+n|+|m_2+m_3+n|
\end{align}

\noindent
The contributions of the vector and chiral multiplets are given by,

\begin{align}
I_V(\{z_l,m_l\},y,n,x)= &\left(1-z_1^{ \pm1}z_2^{\pm1} z_3^{\pm2} x^{|\pm m_1\pm m_2 \pm 2m_3|}\right)
\left(1-z_1^{ \pm1}z_2^{\pm1}  x^{|\pm m_1\pm m_2 |}\right)
\nn\\
&
\left(1-z_1^{\pm2} x^{2|m_1|}\right)\left(1-z_2^{\pm 2} x^{2|m_2|}\right)\left(1-z_3^{\pm 2} x^{2|m_3|}\right)\left(1-y^{\pm 2} x^{2|n|}\right)
,  
\end{align}

\begin{align}
I_\chi ( \{z_l,m_l\},y,n,x,c)= &PE\Bigg[  \left(c+\frac{1}{c} \right) \left( \frac{x^\frac12 -x^\frac32}{1-x^2} \right)  \left(\sum_{i=1}^2 z_i^{\pm1}z_3^{\pm1}y^{\pm1} x^{|\pm m_i\pm m_3 \pm n|}\right) \Bigg]
\end{align}
where we use the fugacity $c$ for the Cartan of the $SU(2)/\mathbb{Z}_2$ acting on the bifundamental.

We expand the index in powers of $x$. One complication of this model is that $\epsilon_0$ is often negative. Nevertheless, the index appears to converge as the Chern-Simons level renders contributions with negative powers of $x$ to be gauge variant and so do not contribute to the index. An unfortunate biproduct of this is that the index computation becomes technically involved as one needs to expand the $PE$ to high orders, even if one only evaluates the full index to a low order in $x$. As such, here we have merely managed to go to order $x^4$, performing the computation for configurations with $|\epsilon_0| \leq 11$. The contributing monopoles are listed in Table \ref{f4tab}

\begin{table}[htpb]
		\centering
		\begin{adjustbox}{center}
			\begin{tabular}{|c|c|c|}
				\hline
				Monopole & Contribution  \\
				\hline
				\multirow{2}{*}{$(0,0,0,0)$} &$1+x+\Big(\chi_{\bf5}(c)-\chi_{\bf3}(c)+2\Big)x^2+\Big(-2\chi_{\bf3}(c)+2\Big)x^3$\\
                &$+\Big(\chi_{\bf9}(c)-\chi_{\bf7}(c)-\chi_{\bf3}(c)+2\Big)x^4...$\\\hline
				\multirow{2}{*}{$(1,0,0,1)$, $\left(\frac{1}{2},\frac{1}{2},\frac{1}{2},1\right)$} & $\chi_{\bf{5}}(c)x^2+\Big(\chi_{\bf{7}}(c)-\chi_{\bf3}(c)-1\Big)x^3$\\
                &$+\Big(\chi_{\bf9}(c)-\chi_{\bf7}(c)-\chi_{\bf5}(c)+2\chi_{\bf3}(c)\Big)x^4+...$\\\hline
				 $(1,0,0,0)$, $\left(\frac{1}{2},\frac{1}{2},\frac{1}{2},0\right)$ & $-x^3+\Big(2\chi_{\bf3}(c)-1\Big)x^4+...$
				\\\hline
				$(1,0,1,2)$, $\left(\frac32,\frac12,\frac12,2\right)$ & $\chi_{\bf9}(c)x^4+...$\\
				\hline
			\end{tabular}
		\end{adjustbox}
		\caption{Monopoles contributing to the $F(4)$ index. The contributions shown include the monopole and all its Weyl equivalents. Summing all contributions reproduce the result \eqref{f4ind}. The result is up to monopoles with $|\epsilon|\leq 11$ and $|m_1|,|m_2|,|m_3|,|n|\leq 15$.}
		\label{f4tab}
	\end{table}  

For the $Spin(7)_{-3} \times SU(2)_2 / \mathbb{Z}_2 $ case the index is the exact same expression but now the allowed monopoles are such that ,
\begin{equation}
    \begin{gathered}
        m_1+m_3+n\in\mathbb{Z},\;\;\;\;\; m_2+m_3+n\in\mathbb{Z},\\
        m_1,m_2,m_3,n\in\mathbb{Z}/2,
    \end{gathered}
\end{equation}
% \begin{align}
% m_1+m_3+&n \in \mathbb{Z},\;\;\;\;\;  m_2+m_3+n \in \mathbb{Z}
% \nn\\
% & m_1,m_2,m_3,n \in \mathbb{Z}/2
% \end{align}
 in particular $n$ can now take half-integer values.  In additional to the monopoles listed in table \ref{f4tab}, we have contributions from $(\frac12,\frac12,0,\frac12)$, $(0,0,\frac12,\frac12)$ and $(1,0,\frac{1}{2},\frac32)$ monopoles which are not allowed in the $F(4)$ case. Their contribution is given in Table \ref{f4z2tab}. Here additional monopole configurations were taken up to $|\epsilon_0|\leq 7$.

\begin{table}[htpb]
		\centering
		\begin{adjustbox}{center}
			\begin{tabular}{|c|c|c|}
				\hline
				Monopole & Contribution  \\
				\hline
				\multirow{2}{*}{$(0,0,\frac{1}{2},\frac{1}{2})$, $(\frac12,\frac12,0,\frac12)$} &$\chi_\mathbf{3}(c)x+\chi_{\mathbf{5}}(c)x^2+\Big(\chi_\mathbf{7}(c)-\chi_{\mathbf{5}}(c)-2\chi_{\mathbf{3}}(c)-1\Big)x^3$\\
                & $+\Big(\chi_{\mathbf{9}}(c)-\chi_{\mathbf{7}}(c)+3\chi_{\mathbf{3}}(c)+2\Big)x^4+...$\\\hline
				$(1,0,\frac12,\frac32)$ & $\chi_\mathbf{7}(c)x^3+\Big(\chi_\mathbf{9}(c)-\chi_\mathbf{5}(c)\Big)x^4+...$\\
				\hline
			\end{tabular}
		\end{adjustbox}
		\caption{Additional monopoles contributing to the $F(4)/\mathbb{Z}_2$ index. The contributions shown include the monopole and all its Weyl equivalents. Summing all contributions reproduce the result \eqref{f4z2ind}. The result is for monopoles with $|\epsilon|\leq 8$ and $|m_1|,|m_2|,|m_3|,|n|\leq 10$ and up to $\mathcal{O}(x^4)$ .}
		\label{f4z2tab}
	\end{table}

\subsection{ $SO(4)_{-2}\times USp(2)_1$ and  $Spin(4)_{-2}\times USp(2)_1$} \label{SOapp}

The index of the  $SO(4)_{-2k}\times USp(2)_{k}$ case is given by,

\begin{align}\label{F4ind}
\mathcal{I}_{SO(4)_{-2k} \times USp(2)_{k} }  = \frac{1}{8} \sum_{\{m_i\},n } \oint \prod_{i=1}^2 \frac{dz_i}{2\pi i z_i}  \frac{dy}{2\pi i y}&z_i^{-2k m_i}y^{2k n}  x^{\epsilon_0} \nn
\\
& I_V(\{z_l,m_l\},y,n,x) I_{\chi} (\{z_l,m_l\}, y,n,x,c)
\end{align}

where $z_i$ parametrize the $SO(2)\times SO(2)$ subgroup of $SO(4)$, and $m_i$ are the corresponding GNO monopole charges. Similarly $y,n$ parametrize the $USp(2)$ part of the gauge group and the corresponding monopole charge. In the chosen basis the summation above is over integer $m_i, n$.

The zero-point energy is given by,

\begin{align}
\epsilon_0(m_1,m_2,n)= &|m_1-n|+|m_1+n|+|m_2-n|+|m_2+n|\nn\\ &-2|n|-|m_1-m_2|-|m_1+m_2|
\end{align}
The contributions of the vector and chiral multiplets are given by,

\begin{align}
I_V(\{z_l,m_l\},y,n,x)= &\left(1-z_1^{ \pm1}z_2^{\pm1}  x^{|\pm m_1\pm m_2|}\right)
\left(1-z_1^{ \pm1}z_2^{\pm1}  x^{|\pm m_1\pm m_2 |}\right)
\left(1-y^{\pm 2} x^{2|n|}\right)
,  
\end{align}

\begin{align}
I_\chi ( \{z_l,m_l\},y,n,x,c)= &PE\Bigg[  \left(c+\frac{1}{c} \right) \left( \frac{x^\frac12 -x^\frac32}{1-x^2} \right) \left( \sum_{i=1}^2 z_i^{\pm1}y^{\pm1} x^{|\pm m_i\pm n|}\right) \Bigg]
\end{align}

 The computation then is straightforward. We take $k=1$ and expand in powers of $x$. Since the zero-point energy here is always positive, the computation is simpler. The contributing monopoles up to order $x^4$ are listed in table \ref{SOtab}. Thus, we get the result given in eq. \eqref{SOind}.

\begin{table}[htpb]
		\centering
		\begin{adjustbox}{center}
			\begin{tabular}{|c|c|c|}
				\hline
				Monopole & Contribution  \\
				\hline
				\multirow{2}{*}{$(0,0,0)$} &$1+x+\Big(\chi_{\bf{5}}(c)-\chi_{\bf3}(c)+2\Big)x^2+\Big(-2\chi_{\bf3}(c)+2\Big)x^3$\\
                &$+\Big(\chi_{\bf9}(c)-\chi_{\bf7}(c)+2\Big)x^4+...$\\\hline
                $(0,0,1)$ & $-x^4+...$\\\hline
                $(1,0,0)$ & $x^2-\chi_{\bf3}(c)x^3+\Big(-\chi_{\bf3}(c)+1\Big)x^4+...$\\\hline
                \multirow{2}{*}{$(1,0,1)$} & $\chi_{\bf3}(c)x+\Big(\chi_{\bf5}(c)-1\Big)x^2+\Big(\chi_{\bf7}(c)-\chi_{\bf5}(c)-\chi_{\bf3}(c)-1\Big)x^3$\\
                &$+\Big(\chi_{\bf9}(c)-\chi_{\bf7}(c)+4\chi_{\bf3}(c)+1\Big)x^4+...$\\\hline
                $(1,1,1)$ & $-2x^3+2\chi_{\bf3}(c)x^4+...$\\\hline
                \multirow{2}{*}{$(2,0,2)$} & $\chi_{\bf5}(c)x^2+\Big(\chi_{\bf7}(c)-\chi_{\bf3}(c)\Big)x^3$\\
                &$+\Big(\chi_{\bf9}(c)-\chi_{\bf7}(c)-\chi_{\bf5}(c)+\chi_{\bf3}(c)\Big)x^4+...$\\\hline
                $(3,0,3)$ & $\chi_{\bf7}(c)x^3+\Big(\chi_{\bf9}(c)-\chi_{\bf5}(c)\Big)x^4+...$\\\hline
                $(4,0,4)$ & $\chi_{\bf9}(c)x^4+...$
				\\\hline
			\end{tabular}
		\end{adjustbox}
		\caption{Monopoles contributing to the $SO(4)_{-2}\times USp(2)_1$ index up to $\mathcal{O}(x^4)$. The contributions shown include the monopole and all its Weyl equivalents. Summing all contributions reproduce the result \eqref{SOind}. The result was checked up to monopoles with $|m_1|,|m_2|,|n|\leq 10$.}
		\label{SOtab}
	\end{table}  
The index for the $Spin(4)_{-2}\times USp(2)_1$ case is very similar, but now the GNO monopole charges satisfy the condition
\begin{align}
    m_1+m_2\in2\mathbb{Z}\,,
\end{align}
which excludes the sectors $(1,0,0)$, $(1,0,1)$, and $(3,0,3)$ from Table \ref{SOtab}.
\bibliographystyle{JHEP}
\bibliography{ref}

\end{document}